\documentclass[twocolumn,showpacs,preprintnumbers,amsmath,amssymb]{revtex4}
\usepackage{graphicx}
\usepackage{dcolumn}
\usepackage{bm}

\begin{document}

\title{Electronic structure of superoxygenated La$_2$NiO$_{4}$ domains with ordered oxygen interstitials}

\author{Thomas Jarlborg$^{1,2}$ and Antonio Bianconi$^{2,3,4}$}

\affiliation{
$^1$DPMC, University of Geneva, 24 Quai Ernest-Ansermet, CH-1211 Geneva 4,
Switzerland
\\
$^2$ RICMASS Rome International Center for Materials Science Superstripes, Via dei Sabelli 119A, 00185 Rome, Italy
\\
$^3$ Institute of Crystallography, Consiglio Nazionale delle Ricerche, via Salaria, 00015 Monterotondo, Italy
\\
$^4$ INSTM, Consorzio Interuniversitario Nazionale per la Scienza e Tecnologia dei Materiali, Udr Rome, Italy}


\begin{abstract}

The electronic structures of La$_2$NiO$_{4+\delta}$, where additional oxygen
 interstitials
 are forming stripes 
along (1,1,0), are presented. Spin-polarized calculations show that ferromagnetism on Ni sites
is reduced near the stripes and enhanced far from the stripes. Totally the magnetic moment becomes
reduced because of oxygen interstitials. It is suggested that the oxygen interstitial concentration in oxygen rich domains in nickelates
suppress magnetism and give multiband metallic domains.

\end{abstract}

\pacs{74.20.Pq,74.72.-h,74.25.Jb}

\maketitle

\section{Introduction.}

Superoxygenated La$_2$NiO$_{4+\delta}$  is an oxygen ionic conductor at high temperature \cite{1,2,3,4,5,6} with 
interesting electrochemical, thermo-mechanical  properties and applications in fuel cells \cite{7,8,9,10}. 
The gas of mobile oxygen interstitials (O$_i$)at high temperate freeze below about 400K forming crystalline grains, as is
observed by electron and neutron diffraction \cite{11,12,13,14,15}. At lower temperature, below 200K,  
the electronic structure show the onset of spin and charge stripes i.e., one dimensional spin density waves (SDW) 
and charge density waves (CDW) \cite{16,17,18,19,20} due to ordering of polarons \cite{21}. 
Localized charges trapped by local lattice distortions (LLD) make periodic lattice distortions (PLD), 
as obtained by doping a magnetic Mott insulator La$_2$NiO$_{4}$. The polaronic CDW in La$_2$NiO$_{4+\delta}$ 
is therefore a model system to be compared with the more complex case in doped cuprates where polarons
\cite{bia1,bia2,bia3} coexist with free carriers \cite{bia4,bia5,bia6}  giving origin to Feshbach like resonances of polaron pairs and BCS pairs
 \cite{bia7,bia8,bia9,bia10,bia11} in the superconducting phase.  The other common feature of nickelates and cuprates 
is the electronic phase separation  in multi-orbital strongly correlated systems which is predicted by the 
multiband Hubbard model in presence of long range Coulomb interaction  \cite{kugel1,kugel2}.
The stripe inhomogeneity of the lattice due to i) ordering of oxygen interstitials in ordered domains \cite{poccia} 
ii) oxygen mobility at high temperature and the electron-lattice interaction at  low temperature in nickelates 
and cuprates is controlled by the lattice mismatch between the  CuO$_{2}$  or NiO$_{2}$  2D atomic layers and the  
La$_2$O$_{2 +\delta}$ blocks layers.  \cite{strain1,strain2} which is key feature for the lattice stripes phase in cuprates \cite{phillips,campi}.

La$_2$NiO$_{4 +\delta}$ is isostructural with the widely studied La$_2$CuO$_{4 +\delta}$ which could make its 
band structure very similar to optimally doped La$_2$CuO$_{4}$ (LCO) \cite{jbmb,jbb}, even if the band 
filling is different. Ordering of excess
oxygens in interstitial positions in LCO will also enhance $T_c$ \cite{poccia}, and the Fermi
surface (FS) is shown to become fragmented by the oxygens \cite{jb}.
Moreover it is similar with other systems where oxygen interstitials like in HgBa$_2$CuO$_{4 +\delta}$ or oxygen 
vacancies like in Ba$_2$CuO$_{4 - \delta}$ 
increases considerably $T_c$  \cite{jin,geba,chma}.
The electronic structure of oxygenated nickelates  La$_2$NiO$_{4+\delta}$ is not known.
For instance it is not established that ferromagnetism (FM) can be suppressed and
allow for superconductivity in these
materials. The equivalence of spin-phonon coupling in the cuprates, where phonons can enforce spin waves \cite{tj7},
is not known. It can lead to fluctuations of local moment amplitudes as in other FM materials \cite{fesi,cevib}.
Therefore it is of high interest to  calculate the electronic structure of the domains present in  La$_2$NiO$_{4+\delta}$  below 400K with a 
stoichiometric content of oxygen interstitials.
In this work we present electronic structure results for the highly hole doped puddles of oxygen interstitials 
in superoxygenated La$_2$NiO$_{4}$, ordered into stripes along (1,1,0) mostly separated by 3 unitcells.
The method of calculation is presented in sect. II.  Experimental information on oxygen ordering  is used
to define the supercells of O-rich LNO, as discussed in sect. II.  
In sect. III we discuss the results of the calculations, and some ideas for future works are given together with the 
conclusions in sect. IV. 

\section{Method of calculation.}

The calculations are made using the linear muffin-tin orbital (LMTO) method \cite{lmto,bdj} and the
local spin-density approximation (LSDA) \cite{lsda}. 
The details of the methods have been published earlier 
\cite{jb},\cite{tj1}-\cite{tj11}. The elementary cell of La$_2$NiO$_4$ (LNO) contains La sites 
at (0,0,$\pm$.721c), Ni at (0,0,0), planar O's at (0.5,0,0) and (0,.5,0) and apical
O's at (0,0,$\pm$.366c), in units of the lattice constant $a_0$=3.86 \AA, where c=1.16.
In addition to the MT-spheres at the atomic sites
we insert MT-spheres at positions (.5,0,$\pm$.5c) and (0,.5,$\pm$.5c) to account for the positions
of empty spheres.

\begin{table}[ht]
\caption{\label{table54}
Decomposition of the total DOS at $E_F$ on each of the six Ni atoms in La$_{12}$Ni$_6$O$_{25}$ (LNO-1)
and La$_{12}$Ni$_6$O$_{24}$ (LNO-0) (in units of $(cell \cdot eV)^{-1}$),
 magnetic moment per site, $m$, ($\mu_B$ per site), and number of valence electrons per Ni site, $Q$. 
 The total magnetic moment in the LNO-0 cell is 1.32 $\mu_B$ and in LNO-1 0.76 $\mu_B$.
 The interstitial site in LNO-1 is at the $y$-layer between site 1 and 2.
  }
  \vskip 2mm
  \begin{center}
  \begin{tabular}{l c c c c c c c c}
  \hline
  ~  & 1 & 2 & 3 & 4 & 5 & 6 & ~ & in LNO-0 \\
  \hline \hline

 N$(E_F)$& 1.2 & 1.6 & 2.4  & 2.6  & 2.4 & 1.6 & ~ & 1.9 \\
      $m$ & 0 & -0.03 & 0.20 & 0.25 & 0.19 & -0.02 & ~ & 0.18 \\
      $Q$  & 9.09 & 9.10 & 9.11 & 9.12 & 9.11 & 9.10 & ~ & 9.15 \\

  \hline
  \end{tabular}
  \end{center}
  \end{table}
  
  \begin{table}[ht]
\caption{\label{table72}
The local decomposition of the DOS at $E_F$ on the different Ni atoms in La$_{16}$Ni$_8$O$_{33}$ (LNO8-1)
and La$_{16}$Ni$_8$O$_{34}$ (LNO8-2). The last column shows the total DOS per cell, all in units of 
$(cell \cdot eV)^{-1}$).
Site 1 is nearest to the oxygen interstitials, site 5 is most distant, and the other sites are counted pairwise
as function of increasing distance from the O-rich layer.  }
  \vskip 2mm
  \begin{center}
  \begin{tabular}{l c c c c c c c }
  \hline
 cell  & 1 & $\pm$2 & $\pm$3 & $\pm$4 & 5 & ~ & total \\
  \hline \hline

LNO8-1 N$(E_F)$& 0.8 & 1.3  & 1.9  & 2.4 & 2.7 & ~ & 25 \\
LNO8-2 N$(E_F)$& 1.3 & 1.0  & 1.3  & 1.8 & 2.3 & ~ & 20 \\

  \hline
  \end{tabular}
  \end{center}
  \end{table}

\begin{figure}
\includegraphics[height=8.0cm,width=9.2cm]{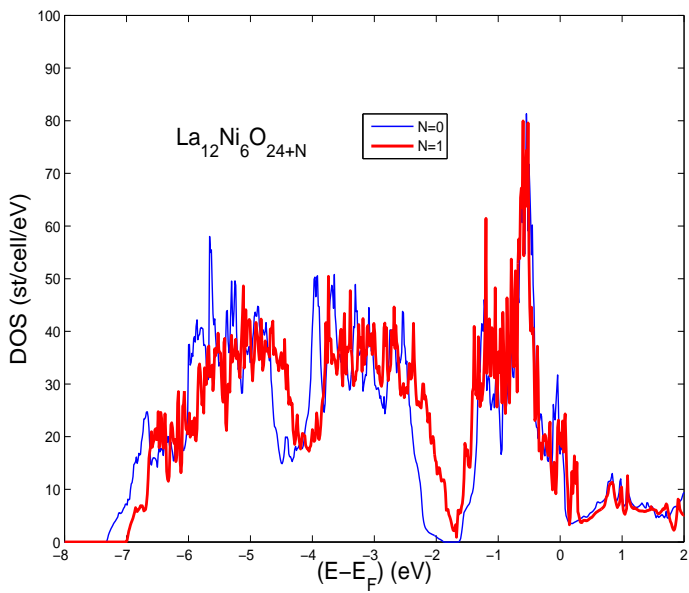}
\caption{(Color online)  The total DOS for La$_{12}$Ni$_6$O$_{24+N}$ with $N$=0 and $N$=1. }
\label{fig1}
\end{figure}

\begin{figure}
\includegraphics[height=8.0cm,width=9.2cm]{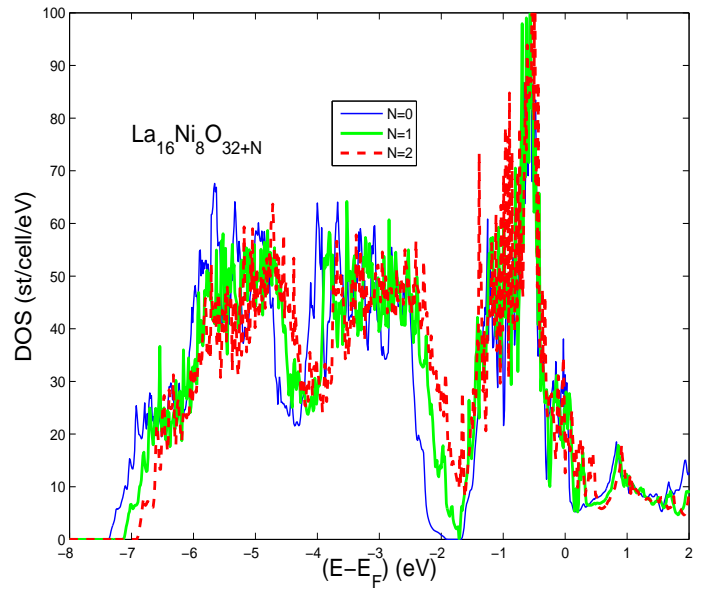}
\caption{(Color online) The total DOS for La$_{16}$Ni$_8$O$_{32+N}$ with $N$=0, 1 and 2. }
\label{fig2}
\end{figure}

The atomic sphere radii are 1.75 \AA~(La), 1.25 \AA~ (Cu), 1.17 \AA~ 
(planar oxygens, oxygen, interstitials, O$_i$, and empty spheres), and 1.20 \AA~ (apical oxygens), respectively.
Six units of the elementary cell La$_2$NiO$_4$ are put together to form a long supercell, La$_{12}$Ni$_6$O$_{24}$,
with the generating
lattice vectors $(1,-1,0), (3,3,0)$, and $(.5,.5,c)$, so that 
its axis is oriented parallel 
to the (1,1,0)-direction, i.e. at 45 degrees
from the Ni-O bond direction along (1,0,0). 
The band calculations are made for this supercell containing 54 sites totally, where one of the empty
sites at the interstitial positions is occupied by the excess oxygen ion. 
These supercells are chosen in order to represent fairly well the experimentally determined structures \cite{frat}.
The basis set goes
up through $\ell$=2 for all sites. The z-projected IBZ (irreducible Brillouin zone) corresponding to the supercell 
is shown
in Fig. \ref{fig3} by the limits $\Gamma-M_3-X_3-R$. It corresponds to one third (folded) BZ
for antiferromagnetic (AFM) LCO. Paramagnetic and spin-polarized calculations are made for these cells.
Self-consistency is made with 192 k-points and final
results are based on 702 points in the IBZ. 

No interstitial sites are occupied with O in
one set of the calculations (called LNO-0). 
The LNO-0 results serve as a reference for comparison with the results with oxygen interstitials.
For instance, the FS for (undoped)
LNO forms a circle centered at the $\Gamma$ point in BZ of the elementary cell, but since the
BZ corresponding to the supercell is folded and very flat along (1,1,0) it is necessary
to identify the circle from several FS pieces in the folded zone. 
In the second set of calculations we insert one oxygen (La$_{12}$Ni$_6$O$_{24+1}$, LNO-1) 
 at an interstitial position to calculate the electronic structure of a La$_2$NiO$_{4.166}$  domain.
 
Paramagnetic calculations were also made for La$_{16}$Ni$_8$O$_{32+N}$ by putting together 8 elementary cells 
of La$_2$NiO$_4$ to calculate the electronic structure of a La$_2$NiO$_{4.125}$  domain.  Totally there are 72 sites in these supercells, and
$N=$0, 1 or 2 interstitial oxygens are inserted (LNO8-0, LNO8-1 and LNO8-2, respectively). These cells are of equivalent size as the cells
that was used for the studies of electronic structures of interstitial O in LCO \cite{jb},
which is helpful for direct comparisons between the nickelates and cuprates.
An important difference between LNO and LCO is that there is one less filled d-band per metal atom.  
In the largest cell of LNO8-0 there are 208 occupied bands, in LCO8-0
there are 212.
Here, for LNO we concentrate the investigations of the shorter cells, since they correspond 
best to the experimentally found periodicity of O-interstitials in LNO  \cite{11,12,13,14,15}.

The excess O$_i$'s sit at the interstitial interlayer positions,
above the oxygen ion in the NiO$_2$ plane of the  
the orthorhombic unit cell.
The insertion of O$_i$'s is expected to induce hole doping, because  
each new oxygen interstitial will bring 4 new bands well below $E_F$ (one "s" and 3 "p"), but the oxygen has 
only 2 "s" and 4 "p" electrons. Therefore, simple arguments suggest that one Ni-O  band becomes unfilled, i.e. 
$E_F$ has to go down relative to the rest of the bands. However,  
other atoms like La serve as charge reservoirs, lattice reconstructions are likely, and in addition the excess O 
positions are ordered in stripe-like patterns like in La$_2$CuO$_{4+\delta}$ \cite{frat}.

Correlation is not expected to be an issue for cuprates and nickelates with hole doping larger than 0.2 holes per Cu (or Ni) site away from half-filling of the d-band. 
This is confirmed for cuprates from ARPES (angular-resolved photoemission
spectroscopy) and ACAR (angular correlation of positron annihilation radiation), which detect FS's and
bands that evolves with doping larger than 0.2 holes per Cu site in agreement with DFT (density-functional theory)
calculations \cite{pick,dama,posi}. 

\section{Results and discussion.}

 The nonmagnetic (NM) total DOS at the Fermi level for La$_{12}$Ni$_6$O$_{24+N}$ 
 and La$_{16}$Ni$_8$O$_{32+N}$ 
is shown in Figs. \ref{fig1}-\ref{fig2}. 
 It can be seen from the partial gap 1.5-2 eV below $E_F$ 
that the band filling decreases when one (or two) oxygen interstitials (O$_i$) are added in form of stripes. 
In contrast to LCO \cite{jb} there is no clear variation of the total $N(E_F)$ as function
of the number of O$_i$. The total $N(E_F)$ in LNO-0 and LNO-1, are comparable; both about
19 $(cell \cdot eV)^{-1}$, and in LNO8-0, LNO8-1 and LNO8-2 about 26, 25 and 20 $(cell \cdot eV)^{-1}$.
Also in contrast to LCO, there is no strikingly high $N(E_F)$ on the interstitial oxygen sites. 
The local DOS on Ni-sites near and far from the interstitial site
are very different as shown in Tables \ref{table54}. The same trends are found for the larger LNO8-cells, 
see Table \ref{table72}. The DOS at $E_F$ on
Ni-sites nearest to the interstitial O are lower than the average, while in the region far from the
interstitials the $N(E_F)$ values are significantly larger than for Ni in LNO-0. 
However, this
local distribution of the Ni-DOS is delicate: at 0.25-0.30 eV above $E_F$ the distribution
is reversed so that local DOS is peaked ($\sim 3 (cell \cdot eV)^{-1}$) nearest to O$_i$, and it 
goes down rapidly for the
next layers, and reaches $\sim 0.5 (cell \cdot eV)^{-1}$ on the most distant Ni site.
The differences in local $N(E_F)$
are also reflected in the local moments on Ni, as can  be expected from the DOS at $E_F$ and the criterion for Stoner
magnetism.  Ferromagnetism (FM) tends to disappear near layers with interstitial O. 
The moment on the Ni close to the O$_i$-site is practically zero,
and in the next Ni layers the moments are even slightly negative. Further away the moments become
clearly FM again, and their amplitude are even larger (0.2-0.25 $\mu_B$)
than the moment per Ni-site in undoped LNO (0.18 $\mu_B$).
The obvious question is whether FM in nickelates is responsible for the absence of superconductivity
in LNO. However, the domains with increased O$_i$-concentrations made of ordered O$_i$-stripes
show weaker FM, but they are still metallic with a different multiband FS from what is found in LCO superconductors.

The main difference between LCO and LNO with oxygen interstitials, is due to the fact
that Ni has one electron less that Cu. Thus, LNO is like a heavily hole doped version of LCO with
$E_F$ pushed down within the high DOS of the 3d-bands. This makes the DOS large in undoped LNO,  sufficiently
large for Stoner magnetism, while this is not the case in LCO \cite{bj}.
The changes of the effective charges (see Table \ref{table54}) show a weak hole doping on Ni when the
number of O$_i$ increases. The same trend is found in the LNO8 results, and it is
in agreement with the behavior for LCO. 

\begin{figure}
\includegraphics[height=8.0cm,width=9.0cm]{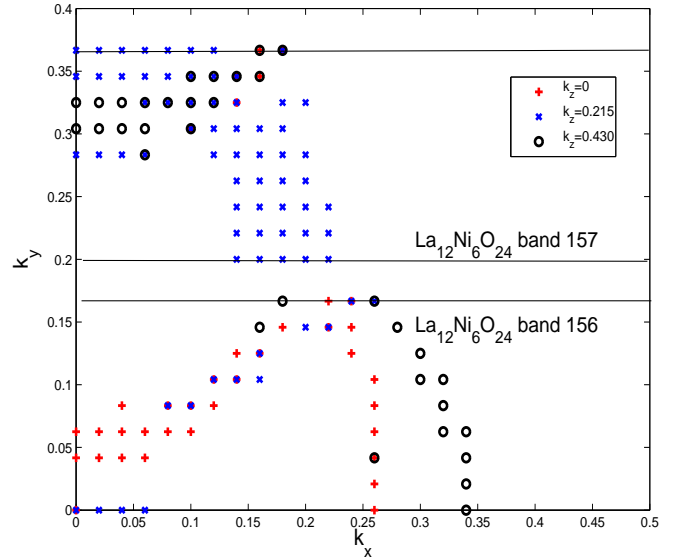}
\caption{(Color online) Upper panel: The Fermi surface of band 157 (upper panel) 
and band 156 (lower panel) for non-magnetic La$_{12}$Ni$_6$O$_{24}$ indicated by 
k-points ($k_x,k_y,k_z$) for which $E(K)$ is within 0.01 eV from $E_F$. The different marks indicate three
different levels of $k_z$. The limits of the rectangular IBZ of the supercell given by the 
rectangle $\Gamma$-M$_3$-X$_3$-R. Unfolding the FS of band 156 leads to circular FS within
the IBZ elementary cell.
}
\label{fig3}
\end{figure}

\begin{figure}
\includegraphics[height=8.0cm,width=9.0cm]{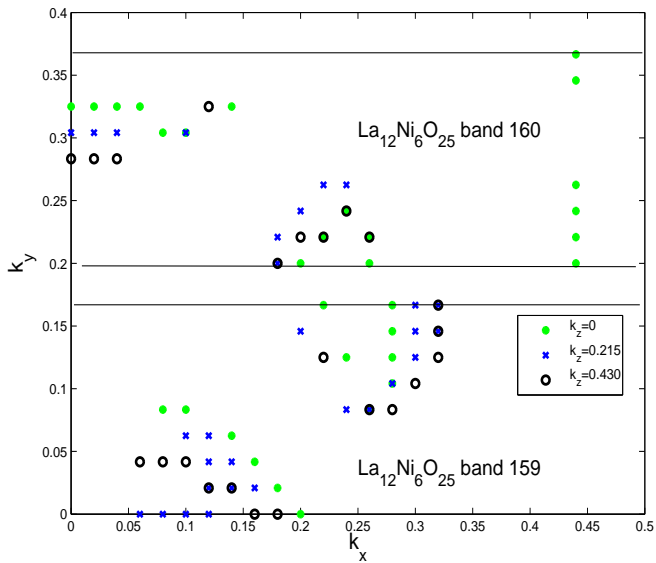}
\caption{(Color online) The FS of La$_{12}$Ni$_6$O$_{24+1}$ displayed as in Fig. \ref{fig3}.
}
\label{fig4}
\end{figure}

In Figs. \ref{fig3}-\ref{fig4} are displayed the FS pieces in different $k_z$-planes of the irreducible BZ
for LNO-0 and LNO-1, respectively.
For the calculations without oxygen interstitial it is easy to recognize simple $\Gamma$-centered FS circles in the 
unfolded BZ, as for LCO \cite{jb}. There are minor modifications in different $k_z$ planes because
of a weak 3-D dispersion. The diameter of the circles is smaller than in undoped LCO. As for stripes in LCO,
the FS's become segmented and show gaps when there are O$_i$ stripes, and it becomes
more difficult to visualize their projection in an unfolded zone. 

\section{Conclusion.}

Oxygen interstitials forming of atomic stripes in the spacer layers form three overlapping mini-bands 
crossing $E_F$ making the local Ni-d DOS larger on sites far from the O$_i$-stripes. Near the stripes the
local Ni-d DOS is reduced. The consequence is that FM is enhanced between the stripes and almost
quenched at the stripes. Thus, the natural growth of O$_i$-stripes in the oxygen rich domains seems to be an efficient way to
 suppress FM in the nickelate, and if the method can be optimized it might be a path for
 making the nickelates good metals. 
It can be noted that suppression of FM is one of the requirements
for superconductivity. The band structures in LNO and LCO are different because of the
different
band filling, but the evolution of the FS's in undoped and oxygenated LNO behave quite
 similarly as in LCO. The undoped material has simple FS's much like the ones for supercells
of stoichiometric LCO despite the large differences of d-band filling between Cu and Ni.
Finally we have shown that  oxygen interstitals order  breaks up the FS's  into fragments with gaps in between, similar to the
process in oxygenated LCO \cite{jb}.

\end{document}